\documentclass[useAMS,usenatbib]{mn2e}
\usepackage{epsfig}
\DeclareGraphicsExtensions{.pdf, .jpg}
\usepackage{bethmacros}

\def\spose#1{\hbox to 0pt{#1\hss}}
\def\lta{\mathrel{\spose{\lower 3pt\hbox{$\mathchar"218$}}
     \raise 2.0pt\hbox{$\mathchar"13C$}}}
\def\gta{\mathrel{\spose{\lower 3pt\hbox{$\mathchar"218$}}
     \raise 2.0pt\hbox{$\mathchar"13E$}}}
\def\kms{{km s$^{-1}$}}

\def\gtrsim{> \sim}
\newcommand\msun{M_\odot}
\def \lya  {Ly$\alpha$}
\def \ihi  {HI}
\def \iovi  {OVI}

\title[MaGICC Halos]{MaGICC Halos: Confronting Simulations with Observations of the Circum-Galactic Medium at z=0}

\author[Stinson et al.]{G.\,S. Stinson$^{1}$\thanks{Email: stinson `at' mpia.de}, C. Brook $^{2,3}$,  
J. Xavier Prochaska$^{4,1}$, Joe Hennawi$^1$, Andrew Pontzen$^5$, 
\newauthor{Sijing Shen$^4$, J. Wadsley$^{6}$, H. M. P. Couchman$^5$, T. Quinn$^{7}$, Andrea V. Macci\`o$^1$,} 
\newauthor{Brad K. Gibson$^{2,8}$}
\vspace*{6pt}\\
$^{1}$Max-Planck-Institut f\"ur Astronomie, K\"onigstuhl 17, 69117, Heidelberg, Germany\\
$^{2}$Jeremiah Horrocks Institute, University of Central Lancashire, Preston PR1 2HE\\
$^{3}$Departamento de F\'{i}s\'{i}ca Te\'{o}rica, Universidad Aut\'{o}noma de Madrid, E-28049 Cantoblanco, Madrid, Spain\\
$^{4}$Department of Astronomy \& Astrophysics, UCO/Lick Observatory, University of California, 1156 High St, Santa Cruz, CA 95064\\
$^{5}$Astrophysics, University of Oxford, Denys Wilkinson Building, Keble Road, Oxford OX13RH\\
$^{6}$Department of Physics and Astronomy, McMaster University, Hamilton, Ontario, L8S 4M1, Canada\\
$^{7}$Astronomy Department, University of Washington, Box 351580, Seattle, WA, 98195-1580\\
$^{8}$Department of Astronomy \& Physics, Saint Mary's University, Halifax, Nova Scotia, B3H 3C3, Canada}

\begin{document}

\maketitle

\begin{abstract}
We explore the circumgalactic medium (CGM) of two simulated star-forming
galaxies with luminosities $L \approx 0.1$ and $1 L^\star$ generated
using the smooth particle hydrodynamic code {\sc GASOLINE}.  
These simulations are part of the Making Galaxies In a Cosmological Context (MaGICC)
program in which the stellar feedback is tuned to match the stellar
mass--halo mass relationship.  For comparison, each galaxy was also simulated using 
a ``lower feedback'' (LF) model which has strength comparable
to other implementations in the literature.  The ``MaGICC feedback''
(MF) model has a higher incidence of massive stars and an
$\approx 2\times$ higher energy input per supernova.  Aside from
the low mass halo using LF, each galaxy exhibits a metal-enriched CGM
that extends to approximately the virial radius.  A significant fraction of this gas has 
been heated in supernova explosions in the disk and subsequently 
ejected into the CGM where it is predicted to give rise to substantial 
\iovi\ absorption.  
The simulations do not yet address the question of what happens to the
\iovi\ when the galaxies stop forming stars.  Our
models also predict a reservoir of cool \ihi\ clouds that show strong
\lya\ absorption to several hundred kpc.  Comparing these models to
recent surveys with the {\it Hubble Space Telescope}, we find that
only the MF models have sufficient \iovi\ and \ihi\ gas in the CGM to
reproduce the observed distributions.  In separate analyses, these
same MF models also show better agreement with other galaxy observables
(e.g.\ rotation curves, surface brightness profiles, and HI gas distribution).  We infer that the CGM
is the dominant reservoir of baryons for galaxy halos.

\end{abstract}

\section{Introduction}

In the paradigm of cold dark matter (CDM) cosmology, gravitational collapse leads to the
formation of virialized and bound  dark
matter halos.  It is predicted that baryons fall into these halos
with the dark matter, comprising a mass fraction up to the
cosmological ratio.  Recent analyses indicates that 
stars comprise less than 25\% of the baryons that should have collapsed into halos \citep{Conroy2009,Mandelbaum2009,Moster2010,Guo2010,More2011}.  
This presents a ``missing baryon'' problem reflected in the unknown whereabouts of the baryons, which
did not cool to form stars.
Substantial amounts of gas are found in a diffuse and
highly ionized medium which is referred to as halo gas or the
circumgalactic medium (CGM).    

The CGM is known to manifest in at least two phases:
(i) a warm/hot, collisionally ionized gas with temperature near
the virial temperature ($T \gtrsim 10^6$K); and 
(ii) a cooler phase (perhaps predominantly photoionized) 
with temperature $T \lta 10^4$K.
The latter phase has been detected in  21cm
emission in our Galaxy in the population
known as the high velocity clouds \citep[e.g.][]{Wakker1997} lying at
distances of tens of kpc \citep{Thom2006}. The hotter phase may be observed via
bremsstrahlung emission with X-ray telescopes, in the intracluster 
medium of massive galaxy clusters \citep[e.g.][]{Allen2001}, but
so far extended diffuse emission has only been detected around one disc
galaxy, the massive NGC1961 \citep{Anderson2011}.  Smaller disc galaxies 
have X-ray bright outflows \citep{Strickland2004}, but halo emission remains undetected.

The diffuse nature of the CGM makes positive detections of
emission rare.  The majority of empirical constraints, therefore, come from
absorption-line analysis of galactic halos, along coincident
sightlines to distant quasars and galaxies
\citep[e.g.][]{Bowen1996,Lanzetta1995,Chen2008,Rubin2010a,Steidel2010}. 
These observations reveal a cool and frequently metal-enriched phase
traced by \ihi\ Lyman series absorption
\citep[e.g.][]{Chen2003} and low-ionization metal-line
transitions \citep[e.g.][]{Chen2008,Barton2009,Rubin2011}.
Importantly, galaxies of essentially all luminosity and spectral type
exhibit significant \ihi\ absorption out to impact parameters $R
\approx 300$\,kpc \citep{Wakker2009,Prochaska2011}, implying a massive ($M
\sim 10^{10} \msun$) extended CGM.  

A particularly useful tracer of the hot phase in quasar absorption line analysis is the \iovi\ doublet, which occurs in highly ionized and enriched regions of the Universe.  If
collisionally ionized, the gas has temperature $T \sim 10^5-10^6$\,K.
\iovi\ is observed along the majority of sightlines through our
Galactic halo and is believed to trace coronal material
on scales of 10--100kpc \citep{Sembach2003}.  This gas is also
associated with the CGM of local $L>0.1L^\star$ galaxies  \citep{Stocke2006,Wakker2009}.  \cite{Prochaska2011} have further
demonstrated that the extended CGM ($R \sim 300$\,kpc) of 
$L \gtrsim 0.1 L^\star$ galaxies has a high covering fraction to \iovi\ and can
account for all of the \iovi\ detected in the present-day universe.
Most recently, \cite{Tumlinson2011} report a nearly 100\%\
incidence of strong \iovi\ absorption for the halos of $L \sim
L^\star$ star-forming galaxies, with an ionized metal mass that likely exceeds
that of the galaxies' interstellar media, demonstrating
that the CGM is a major reservoir of highly ionized metals at $z \sim
0$.  

The observations reveal a multi-phase, highly ionized and
metal-enriched CGM around present-day galaxies \citep[also see][]{Thom2008,Savage2011,Savage2011b,Wakker2012}. The high degree of enrichment
demands that a large mass of gas (and metals) is transported from galaxies and/or their progenitors  to the CGM.  Thus, these observations provide
robust constraints on the processes of gas accretion and feedback
which may be compared directly against models of galaxy formation.

Previous theoretical work on the CGM for individual galaxies has largely focused
on analytic or semi-analytic treatments of idealized gas and dark
matter/temperature profiles \citep{Mo1996,Telfer2002,Maller2004,Binney2009}.  Although
these works provide crucial insight into the nature of the CGM, they
lack proper cosmological context (e.g.\ mergers) and have
generally not included the role of feedback from star-formation in the
galaxy and its satellites.  The latter is now considered critical
 to match a wide-range of galaxy observables including the luminosity function,
the stellar mass--halo mass relationship, and the baryonic Tully--Fisher relationship \citep{Guo2011}. 
Galactic-scale winds likely resulting from stellar feedback are a nearly generic feature of
star forming galaxies \citep{Shapley2003,Weiner2009,Rubin2010} and may
significantly influence properties of the CGM.

A number of groups have now used cosmological simulations to examine
the nature and enrichment of the IGM at $z \sim 0$ and offer
comparisons to quasar absorption line observations \citep[e.g.][]{Cen2006,Dav'e2010,Oppenheimer2011,shs+11,Cen2012}.
These studies have not explicitly examined the physical nature of the
CGM for individual galaxies and likely have insufficient resolution to
perform such analysis.  A few authors have examined the distribution
of gas within individual galactic halos, but have not yet treated the
distribution of metals or the impact of feedback \citep{kh09,stewart11}.
At $z \sim 3$, 
\cite{Kawata2007} have analyzed numerical simulations of outflows in
L$^\star$ progenitors, and found that \iovi\  better reflects the strength
of galactic winds than \ihi.  Most recently, \cite{fpk+11} and  
\cite{Shen2011} have examined the metal enrichment of the CGM with the
latter demonstrating that metals can be ejected to large distances at
early times from L$^\star$ progenitor galaxies while their gravitational potential remains low.

We have recently begun the MaGICC project to use sufficient stellar feedback to
simulate galaxies that match the stellar mass--halo mass relationship.
Early results of the project have shown that  
the ejection of low angular momentum gas via outflows \citep{Brook2011a} 
redistributes angular momentum via large scale galactic fountains
\citep{Brook2011b} and thus play a crucial role in forming disc
galaxies, particularly those without classical bulges.  In this
way, we form galaxies that match scaling
relations between rotation velocity, size, luminosity, colour, stellar
mass,  halo mass, HI mass, baryonic mass and metallicity (Brook \etal
2012, submitted).  \citet{Macci`o2012} also showed that these galaxies have
cored dark matter density profiles.  Here, we test the baryon cycle of these
simulations with the constraints provided by observations of the CGM,
and in particular the column densities of H\,I and O\,VI as observed
at $z=0$.

\section{Simulations}
\label{sec:sims}

We resimulate a suite of simulations drawn
from the McMaster Unbiased Galaxy Simulations (MUGS, \citealt{Stinson2010}).  
The simulations are listed in Table 1.  The highest mass galaxy, HM, is g5664 from MUGS and is about half the Milky Way's halo mass.  The low mass galaxies, LM\_MF and LM\_LF, are rescaled versions of MUGS initial conditions, each a factor of eight lower mass 
than the MUGS simulations, allowing us to explore mass dependence. LM\_MF was also used in \citet{Brook2011b}.
\renewcommand{\thefootnote}{\alph{footnote}}
\begin{table}
\label{tab:data}
\begin{minipage}{90mm}
\caption{Simulation data}
\begin{tabular}{lllllll}
\hline
Name & M$_{tot}$\footnotemark[1] & 
M$_\star$\footnotemark[2] & Luminosity\footnotemark[3] & c$\star$\footnotemark[4]   &
$\epsilon_{rp}$\footnotemark[5] & 
IMF\footnotemark[6] \\
 &
(M$_\odot$) &
(M$_\odot$) &
(L$^\star$)  &
 & 
 & \\
\hline
HM\_MF & $7\times 10^{11}$ & $1.4\times10^{10}$ & 0.84 & 0.1 & 0.175 & C\\
HM\_LF & $7\times 10^{11}$ & $4.8\times10^{10}$ & 0.79 & 0.05 &-- & K\\
LM\_MF & $1.8\times10^{11}$ & $3.7\times10^9$ & 0.15 & 0.05 & 0.1 & C\\
LM\_LF & $8.8\times10^{10}$ & $8.7\times10^9$ & 0.13 & 0.05 & -- & K\\
\hline
\end{tabular}
\footnotetext[1]{M$_{tot}$ is the virial mass of the halo including dark and baryonic matter.}
\footnotetext[2]{M$_\star$ is the total stellar mass.  }
\footnotetext[3]{V-band luminosity compared to $M_V =-21$.  }
\footnotetext[4]{Star forming efficiency.}
\footnotetext[5]{Radiation Pressure Efficiency.}
\footnotetext[6]{Initial Mass Function: C=\citet{Chabrier2003}; K=\citet{Kroupa1993}}
\end{minipage}
\end{table}

The simulations were  evolved using the smoothed particle hydrodynamics (SPH) code \textsc{gasoline} \citep{Wadsley2004}. 
Supernova energy is implemented using the blastwave formalism \citep{Stinson2006}. Metals are ejected from type II supernovae (SNII), type Ia supernovae (SNIa), and the stellar winds of asymptotic giant branch (AGB) stars. Ejected mass and metals are distributed to the nearest neighbour gas particles using the smoothing kernel \citep{Stinson2006}. Metal diffusion is included and metal cooling is calculated based on the diffused metals \citep{Shen2010}.

Two different star formation and  feedback models are employed. The original MUGS simulations formed stars when gas reached a density of 1.0 cm$^{-3}$, used a  Kroupa \etal (1993) IMF   and deposited 0.4$\times$ 10$^{51}$ ergs per supernova  explosion.  We refer to this as the ``lower feedback model" (LF) in this study, but note that this feedback strength is comparable to or even stronger than most implementations that are currently run in the literature \citep{Scannapieco2011}. 
In our ``MaGICC feedback model" (MF),  four changes have been made to our implementation of star formation and feedback:
\begin{itemize}
 \item we use the more common Chabrier (2003) IMF which creates more massive stars for a given stellar mass; 
\item the star formation density threshold is increased to 9.3 cm$^{-3}$;
\item the energy input from supernovae is increased to $10^{51}$ ergs;
\item we include energy from radiation released by the massive young stars before they
explode as supernovae.  
\end{itemize}

Radiation pressure from massive stars can have significant effects on the scales that are resolved in our simulations \citep{Nath2009,Murray2011}. Massive stars typically produce 10$^{50}$ ergs of energy per $\msun$, yet this couples only weakly to the ISM \citep{Freyer2006}. To mimic this inefficiency, we inject a fraction of the energy as thermal energy in the surrounding gas but do {\it not} turn off cooling \citep[for details see][]{Brook2011b}.  Such thermal energy injection is highly inefficient at the spatial and temporal resolution of cosmological simulations, and is rapidly radiated away \citep{Katz92}
This feedback is even less efficient at low resolution.  Following a parameter search designed to match the stellar mass--halo mass resolution from halo abundance matching \citep{Moster2010},  we inject 17.5\% of radiation pressure to the surrounding  gas as thermal energy in the lower resolution (HM) simulation, but only 10\% in the  higher resolution runs(LM).  The overall coupling of energy to the ISM is minimal, but sufficient to reduce star formation in the region immediately surrounding a recently formed star particle. 

The HM\_LF and HM\_MF simulations use the same initial conditions and each simulated galaxy has
an absolute $V$-band magnitude $M_{\rm V} \approx -20.8$ implying a
luminosity $L \approx 0.8L^\star$.  This is somewhat surprising given that
the two runs yield very different stellar masses for the galaxy;  this
difference is compensated by the fact that each galaxy follows a very
different star formation history (see Figure \ref{fig:massev}) such that the different stellar
ages result in comparable V-band luminosity, but $B-V$ colours of 0.37 for HM\_MF and 0.55 for HM\_LF.  HM\_MF was analysed in \citet{Macci`o2012}. 
LM\_LF and LM\_MF use different initial conditions of similar halo mass.  
They are selected because each has
$M_{\rm V}\approx -19$.  The LM\_LF initial conditions run with MaGICC feedback
have $M_{\rm V}\approx -16.2$.
To summarize the internal properties of the galaxies, the lower feedback galaxies suffer from the problems of angular momentum loss that have long plagued galaxy formation simulations: dense central stellar bulges, centrally peaked rotation curves, dark matter cusps and too many stars relative to halo mass compared to observations.  The MaGICC feedback simulations result in galaxies which match the stellar mass--halo mass relation, have slowly rising rotation curves, and dark matter cores. The MaGICC feedback simulations provide significantly better matches to the internal properties of observed disc galaxies.

\section{Results}
\begin{figure}
\resizebox{9cm}{!}{\includegraphics{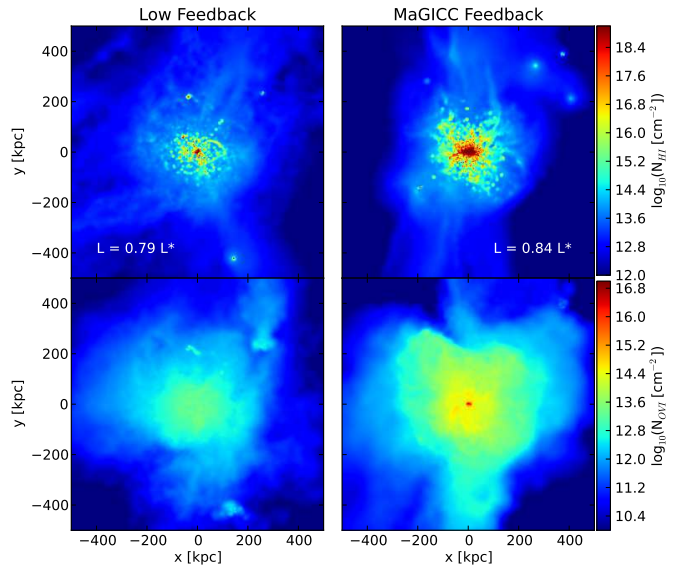}}
 \caption[HI  Maps]{ Column density maps of HI (\emph{top}) and \iovi\ (\emph{bottom}) of the low (\emph{left}) and high (\emph{right}) feedback simulations.  The maps are all aligned so that the discs of the galaxies are edge on to the viewer. The results on scales greater than 10 kpc are relatively independent of viewing angle.}
\label{fig:4map} 
\end{figure}
\begin{figure}
\resizebox{9cm}{!}{\includegraphics{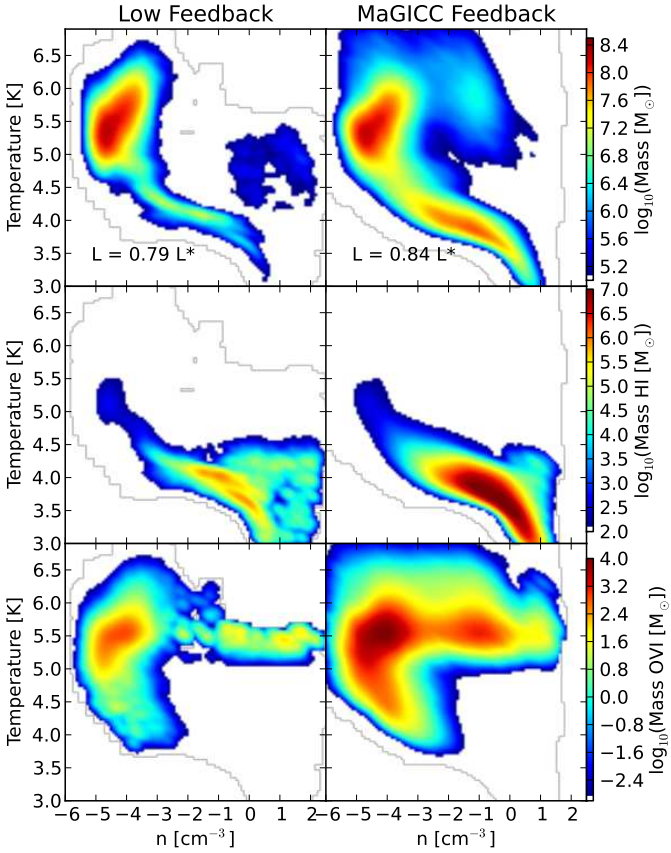}}
 \caption[Phase Diagrams]{ Temperature-density phase diagrams of the galaxy halo simulated with the two different amounts of feedback.  The top plots are mass weighted by the total gas mass, the middle is weighted by the HI mass, and the bottom is weighted by O\,VI.  These phase diagrams only include halo gas, that is gas more than 3 kpc above or below the midplane or at R$_{xy}> 40$ kpc. The distinct locations of OVI and HI show that the hot and cold gas is in different phases in our model.} 
\label{fig:phase} 
\end{figure}

To study the CGM properties of the simulated galaxies, we find the 
galaxy using the Amiga Halo Finder \citep{Knollmann2009} 
and define its center and systemic
velocity by the position and velocity of the particle with the lowest potential.  
We may then sample the CGM at a range of impact parameters $\rho$. 
To generate surface density maps of the CGM, we sum 
each box over 1~Mpc along the line of sight.  The maximum velocity of the material
in the box is $|\delta v| < 200$\,\kms.  To estimate the specific column
densities of \ihi\ and \iovi, we must estimate the
ionization state of the gas along each sightline.  Under equilibrium
conditions, which we assume apply, the ionization state is determined
by the incident radiation field (photoionization) and temperature of
the gas (collisional ionization).  A proper handling of radiative
transfer effects are beyond the scope of this present paper.  Instead,
we examine the simulations in regions expected to
correspond to optically thin material, i.e.\ where the \ihi\ 
column density is low because the gas is expected to be
highly ionized.

We calculated the ionization states for hydrogen
and oxygen throughout the high resolution region assuming optically thin conditions
and the \cite{Haardt2011} UV radiation field evaluated at $z=0$.  With the
Cloudy software package (v10.0 last described in \citet{Ferland1998}), we generated a suite of models
varying the density, temperature, and metallicity of the medium and
used the output to find the OVI and HI fractions for all the gas in the simulation.

In Figure~\ref{fig:4map} we present column density maps for \ihi\ and
\iovi\ for the LF and MF runs of galaxy HM.  Both simulations predict a
CGM extending to at least 100 kpc traced by cool \ihi\ gas and the
more highly ionized \iovi\ gas.  
Figure~\ref{fig:phase} presents three
phase-diagrams of the CGM material in each galaxy.  The top phase diagrams
include the total halo gas mass in the simulations.  The middle only counts
HI mass and the bottom only counts the
mass of the halo OVI.  One notes two distinct phases in the top panels: (i)
a cool ($T \sim 10^4$K), dense ($n > 10^{-2}$ cm$^{-3}$) gas which 
dominates the \ihi\ absorption and 
(ii) a warm/hot ($T > 10^5$K) gas, which the bottom panels show creates the
OVI, primarily through collisional ionization.  

The MaGICC Feedback simulation shows a high density hot component, which
seems unphysical that accounts for 9\% of the halo \iovi\ at $z=0$.  
We find that 90\% of this high density, 
hot halo gas is a relic of the thermal feedback implementation--the gas has had its cooling temporarily disabled--and that this gas 
is concentrated in a $\sim$10 kpc sphere around the galaxy.
This dense, hot gas drives the 
outflows that populate the CGM with metal enriched material, but
it represents a negligible fraction of the \iovi\ column.  We
discuss in \S \ref{sec:xrays} how this dense, hot gas generates
a comparable amount of soft X-ray emission to that observed.

\begin{figure}
\resizebox{9cm}{!}{\includegraphics{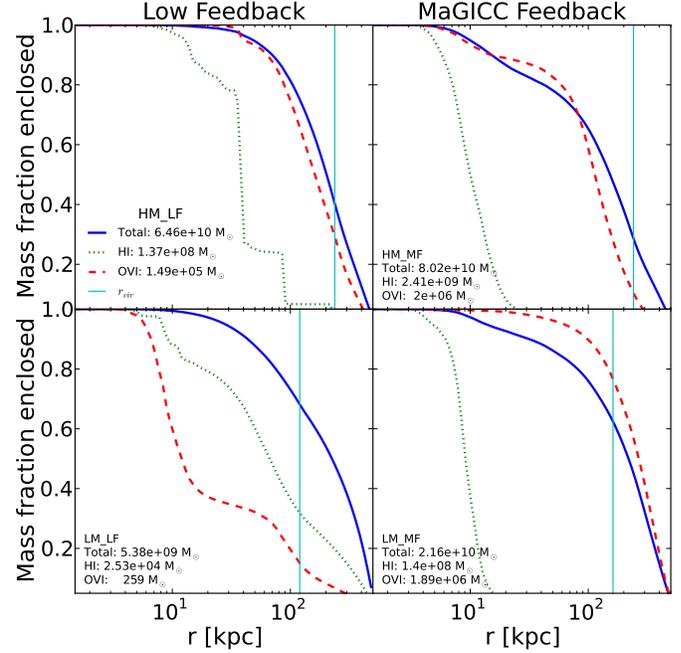}}
 \caption[Gas Mass Profiles]{ Cumulative gas mass profiles from 500 kpc 
inwards for each of the 4 galaxies.  The solid line represents total 
gas mass, the dashed line represents OVI, and the dotted line represent 
HI.  The total mass of each component is given in the legend of each plot.  
With the MaGICC feedback, the HI has a more centrally concentrated 
profile than the OVI. In the $\sim$ L$^\star$ galaxy, HM\_MF, the OVI 
distribution follows the total gas distribution fairly closely, but in 
the lower mass galaxy, the OVI is significantly more extended than the 
total gas, indicating that enriched material is blown farther away.  }
\label{fig:gasmassprof} 
\end{figure}
In terms of total mass, Figure \ref{fig:gasmassprof} shows that the CGM of the HM\_MF galaxy has $4 \times
10^{10} \, \msun$ ($7 \times 10^{10} \, \msun$) of gas to $\rho = 150
(300)$\, kpc.  (We quote 300 kpc since it is the maximum extent at which OVI maintains a 100\% covering fraction in the simulations.)  About half of this CGM has $T < 10^5$\,K giving a
cool gas mass that matches very well with previous empirical estimates
\citep{Prochaska2011}.  Regarding metals, the CGM has $\sim 10^6
\msun$ of \iovi\ to $R = 150$\,kpc, again in excellent
agreement with the mass estimates for $L \approx L^\star$, star-forming
galaxies \citep{Tumlinson2011}.  The LM\_MF model has
a lower CGM mass in both gas and metals by about a factor of 3.  The LM\_LF has 
even less ($\sim 10^9 \msun$) cool gas in the CGM
and a negligible mass of \iovi.  This model is a very poor match to
the observational estimates.

Returning to Figure~\ref{fig:4map}, one notes
that the \iovi\ gas has both a smoother and more extended
distribution than the \ihi.  Again, this reflects the fact that the \iovi\ gas
traces a hotter and more diffuse phase in the halos of this galaxy.
In contrast, the discrete clouds of HI are clustered more closely to the disc and 
are the remnant of both gas-rich mergers and cooling out of the hot halo.
\citet{Maller2004} and \citet{Kaufmann2009} describe how gas can cool out of the hot halo.  
A detailed examination of the evolution of these cold clouds is
beyond the scope of this paper, but a preliminary investigation suggests that the clouds
both cool out of the hot halo and are entrained in the outflows.  The cooling mechanism 
may be a numerical
artifact of SPH whereby particles approach each other due to random
excursions and thus create a slightly higher density, which makes
them cool faster.  Fewer condensations are seen using hydrodynamics
based on the Eulerian (grid) method \citep{Teyssier2002, Agertz2011}
or the new moving mesh code AREPO \citep{Springel2010,
  Vogelsberger2011}.  It remains for future simulations 
 using more detailed radiative transfer to see if
such condensations are an artifact of the SPH hydrodynamics scheme.

\begin{figure}
\resizebox{8cm}{!}{\includegraphics{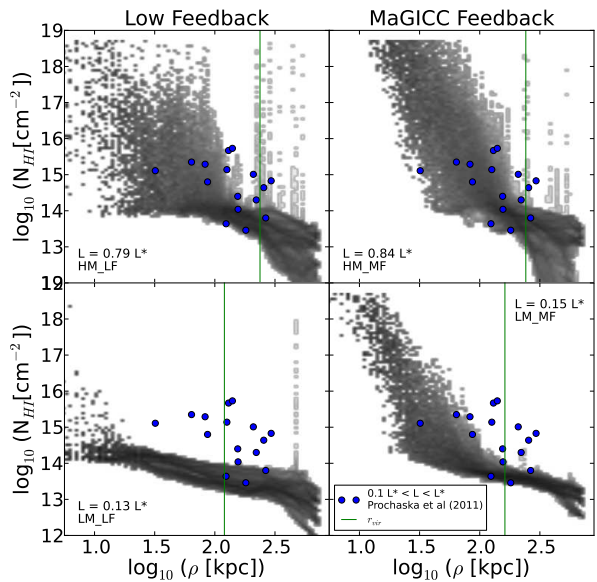}}
 \caption[ Four Galaxy HI Radial Profiles]{ Radial profiles of the column density maps of HI for the four simulated galaxies.  Large blue dots are observations of  0.1 $L^\star<L<L^\star$ from the \citet{Prochaska2011}  galaxy sample while large green squares are  galaxies with L$<$0.1L$^\star$.   
  The solid green line represents the virial radius, $r_{vir}$, for each of the four halos. 
}
\label{fig:6galhiradprof} 
\end{figure}

\begin{figure}
\resizebox{8cm}{!}{\includegraphics{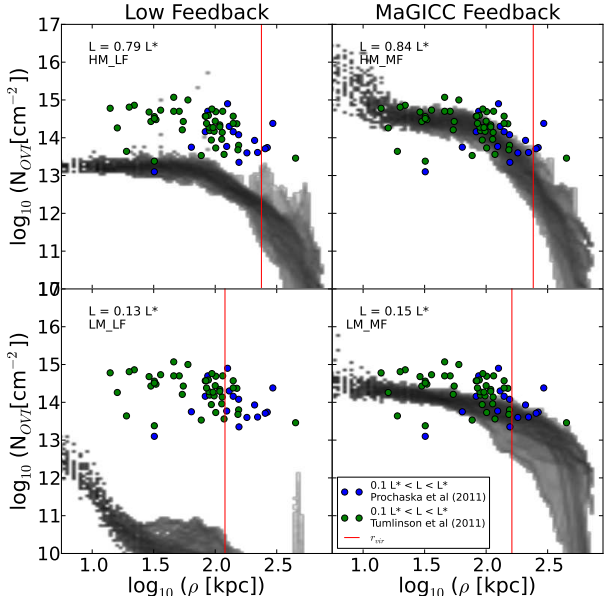}}
 \caption[ Four Galaxy OVI Radial Profiles]{ Radial profiles of the column density maps of OVI for four galaxies spanning a range of masses simulated using MaGICC feedback simulations.  The large dots are observations of  0.1 L$_*$$<$L$<$L galaxies from the \citet{Prochaska2011} (\emph{blue}) and \citet{Tumlinson2011} (\emph{green}) galaxy samples.   
  The solid red line represents the virial radius, $r_{vir}$, for each of the four halos.  }
\label{fig:6galoviradprof} 
\end{figure}

In Figure~\ref{fig:6galhiradprof}, we present the surface density
profiles of \ihi\ gas and in Figure~\ref{fig:6galoviradprof} we present the surface density profiles of \iovi\ gas as a function of the impact parameter
$\rho$ to the center of the galaxy.  Overplotted on the distributions
are observed \ihi\ and \iovi\ column densities for sub-$L^\star$ galaxies
at $z\sim 0$ from \cite{Prochaska2011} and for star-forming $L^\star$
galaxies at $z \sim 0.2$ from \cite{Tumlinson2011}.  We also mark each panel with the 
V-band luminosity of the simulated galaxies and the virial radius, indicated 
by the vertical green line.  Considering first the lower feedback HM\_LF, we find fair agreement 
for the \ihi\ gas although the data
are systematically higher than the majority of simulated sightlines.
The results for \iovi\ are more discordant;  the CGM of HM\_LF
underpredicts the observed \iovi\ column densities by
nearly an order of magnitude at all impact parameters.  In contrast, the HM\_MF predicts nearly
10$\times$ higher \iovi\ surface densities and therefore provides a
reasonable match to the observations.  Furthermore, this model yields
qualitatively better agreement with the \ihi\ observations.  In these
respects, this single simulation of an $L \approx 0.8L^\star$ galaxy has a
CGM with characteristics matching current observations.  At lower masses,
the differences between the high and lower feedback models become starker.  
LM\_LF has an undetectable \iovi\ content beyond 
the optical extent of the galaxy even though the data for galaxies of comparable
luminosity all show detections of \iovi.  In contrast, LM\_MF predicts
a factor of $10^4$ higher \iovi\ surface densities, matching the
observations. 

\subsection{Halo Oxygen Distribution Function}
\begin{figure}
\resizebox{8cm}{!}{\includegraphics{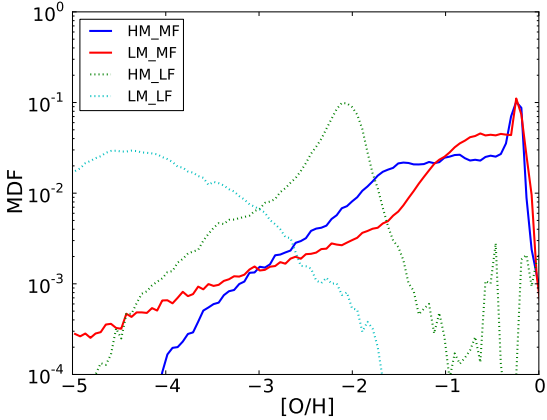}}
 \caption[Halo Gas Oxygen Content]{ The oxygen distribution function for the four galaxies in our sample.  The MaGICC feedback produces a significantly metal-enriched gas halo with a peak in both the HM and LM galaxy just above half solar. }
\label{fig:oxhist} 
\end{figure}
Figure \ref{fig:oxhist} shows the oxygen distribution function of the halo gas in the four simulations.  The MaGICC feedback cases result in a broad distribution of oxygen abundance with a peak at half the solar oxygen abundance comparable to the abundances measured in Milky Way halo gas \citep{Gibson2000,Gibson2001}.  The Lower Feedback simulations are at least 2 orders of magnitude less oxygen enriched.  In total, HM\_MF has 1.2$\times 10^8$ $\msun$ of oxygen in its halo, roughly twice the $7.7\times 10^7$ $\msun$ of oxygen contained in the ISM in the disk, while LM\_MF has roughly an equipartition between the $1.6\times10^7$ $\msun$ in its halo and $2.1\times10^7$ $\msun$ in its disk
\citep[cf][]{Tumlinson2011}.  Lower feedback generates gaseous halos with significantly less oxygen. 

\subsection{Evolution}
\begin{figure}
\resizebox{8cm}{!}{\includegraphics{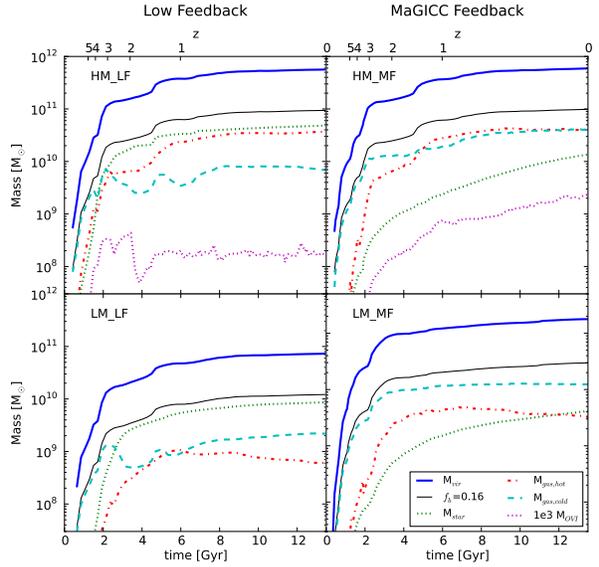}}
 \caption[Mass Evolution]{ The mass evolution of  the total halo mass ({\it blue}), stellar mass ({\it dotted-green}), $10^3\times$\iovi\, mass ({\it dotted-magenta}), total gas ({\it dashed light blue}), and hot (T$> 10^5$ K) gas within $r_{vir}$ for the 4 simulations.  The thin
black line shows the cosmic baryon fraction of the total halo mass.}
\label{fig:massev} 
\end{figure}
To gain a sense of how the CGM developed, Figure \ref{fig:massev} shows the evolution of total mass, along 
with stellar, \iovi\,, cold ($T<10^5$ K), and hot gas 
($T>10^5$ K) masses within $r_{vir}$ for the simulated galaxies.  
The total halo mass increases by two orders of magnitude in the 2 Gyr between $z=6$ 
and 3 in each simulation.  Subsequent merger/accretion events are reflected in mass 
``jumps".  
With lower feedback, most star formation happens early, at the same time the total mass is rapidly increasing.  High feedback delays this star formation.  

In each galaxy, hot gas
develops at the same time as star formation, indicating that stellar feedback
is the initial source of hot gas in galaxy halos.  As the halo grows, accreting cold gas 
is shocked to the virial temperature, so that by $z=0$, only one-third of $T > 10^5$ K gas 
at $z=0$ was directly involved in a stellar feedback event.  The mass history of the \iovi\ 
follows stellar mass in HM\_MF indicating that oxygen produced in supernovae
readily makes its way into the halo as a hot gas. In HM\_LF, the \iovi\ mass initially 
rises, but then drops during
its major merger at $z=1.5$ after which the \iovi\ mass remains constant indicating
that the lower feedback is unable to drive metal-enriched outflows.

It is surprising that even though the feedback implementation is thermal, the MaGICC feedback does not lead
to more hot gas, but rather more cold gas.  As described earlier, this is because
some cold gas gets entrained in outflows and there is rapid cooling of halo gas.
We speculate that the similar amount of hot gas in HM\_LF and HM\_MF that have different feedback implementations is representative of some maximum amount of hot gas that can exist in a halo before the gas cools; a larger parameter search is required to address this question.

\subsection{X-ray Emission}
\label{sec:xrays}
One consequence of predicting that the \iovi\ comes from a massive 
halo of collisionally excited gas
ranging up to $10^7$ K is that it should produce diffuse soft (0.2-2.0 keV) X-ray emission.
Observations of disc galaxies have detected X-rays from outflows less than 10 kpc from the disc, but
never $L_X > 5.2\times10^{39}$ erg s$^{-1}$ outside 10 kpc \citep{Strickland2004,Anderson2011}.
Using a naive \citet{Navarro1995} estimate of $L_X$ based 
solely on bremsstrahlung cooling, HM\_MF has $L_X=5.2\times10^{39}$ ($4.4\times10^{37}$)
erg s$^{-1}$ outside 10 (20) kpc in the 0.2-2.0 keV energy range.  Even with the
large amount of mass in the halo, the gas remains too diffuse to emit significant
X-rays. It is intriguing
that the X-ray luminosity from inside 10 kpc, $L_X=5.8\times10^{40}$, is similar to the observed
diffuse, soft X-ray emission for an $L_B=2.4\times10^{10}$ L$_\odot$
galaxy \citep[e.g.][]{Strickland2004}, but a detailed study of this emission is beyond the scope of this work.

\section{Conclusions}
We have studied CGM gas in high resolution galaxy formation
simulations.  The stellar feedback is constrained using the stellar mass--halo
mass relationship.  This constraint leads to a stellar feedback implementation
that decreases star formation and ejects hot, metal-enriched gas
into the CGM.  Inside the galaxies, the stellar
feedback results in slowly rising rotation curves, flattened dark
matter density profiles and HI
mass-Luminosity, baryonic Tully--Fisher relations, and mass-metallicity
relations that resemble observed galaxies \citep{Brook2012}.  We compared these galaxy models to one with more
standard feedback implementation \citep{Stinson2010, Scannapieco2011}. The lower feedback simulations match
none of these commonly observed internal properties of galaxies.  An examination of
the CGM in our model reveals that: 

\begin{itemize}
\item The MaGICC feedback implementation produces extended, metal-enriched 
gaseous coronae that extend to impact parameters of $\sim$300 kpc, even 
around low mass (L$_*\sim0.1$L$_\odot$) halos.  These results match observed 
absorption line features \citep{Wakker2009,Prochaska2011, Tumlinson2011}.
\item The hot coronae extend beyond $r_{vir}$, though most of their mass remains 
contained within $r_{vir}$ of the simulated galaxies. The coronae extend furthest 
outside $r_{vir}$ in low mass systems.
\item  A reasonably strong feedback implementation, compared with 
those in the literature (our ``lower feedback'' case), is insufficient to 
match \ihi\ absorption features at $z=0$ and severely underestimates \iovi\,,
performing particularly poorly in low mass ($\sim0.1$ L$^\star$) simulations.
\item The amount of hot gas in the LF and MF runs did not differ significantly, but rather: 
\begin{itemize}
\item (i) The MaGICC feedback ejected ten times more oxygen out of the disc than the lower feedback
in the high mass case and a factor of $10^4$ more in the low mass case.
\item (ii) The MaGICC feedback halo contains ten times the cold gas as the 
lower feedback case in both high and low mass simulations.  The cold gas 
results from cooling out of the hot halo gas and cold gas entrained in outflows.
\end{itemize}
\item The MaGICC feedback simulations match the observed \ihi\ and \iovi\ 
absorption columns at a range of masses.
\item The total mass of the coronae is several times higher than the stellar mass 
in the simulated MaGICC feedback galaxies.  Thus, the simulations of individual 
galaxies predict that the ``missing baryons'' are found in the CGM of galaxies.  
\end{itemize}

In our simulations, we only considered galaxies defined as star forming by
\citet{Prochaska2011} and \citet{Tumlinson2011}.  One of the
key results of those works is the absence of OVI detection in quiescent
galaxies.  Our simulations are unable to address the
question of what happens to the OVI when galaxies stop forming stars.

Our examination focused on the radial distribution of \ihi\ and
\iovi\ gas in the CGM of a simulated $L \approx L^\star$ galaxy and
compared these results to observed surface density profiles.  There
are, of course, additional observables related to the entire
population of \iovi\ absorbers identified in quasar absorption line
studies \citep[e.g.][]{Danforth2008,Thom2008a,Tripp2008};  any truly successful
model must also reproduce these resuts. 
One key constraint is the observed distribution of line-widths
(a.k.a. Doppler parameters or $b$-values) for the \iovi\ gas
\citep{Thom2008a,Tripp2008} which gauge, in a statistical fashion, the
contributions of thermal broadening, turbulence, and gas dynamics to
the motions of the gas.  On average, the $b$-values for \iovi\
absorbers are substantial ($b > 20$\kms) and appear to correlate with
the strength of \iovi\ absorption.  This has posed a considerable
challenge to models where \iovi\ is predominantly photoionized
\citep[e.g.][]{od09}, as such gas has a small thermal width.  Another
valuable kinematic constraint is the common observation of co-aligned
\ihi\ and \iovi\ absorption relative to a common redshift
\citep[approximately half of observed \iovi\ systems;][]{Thom2008}. 
Indeed, this appears to be at odds with the apparent separation of
strong \ihi\ absorption (at smaller radii) with the extended \iovi\
absorption of our modeled CGM.  On the other hand, we do observe
non-negligible \ihi\ absorption to large radii (Figure~\ref{fig:6galhiradprof})
which may trace the \iovi\ gas.  A proper comparison of our model with
these observables will require the generation and detailed analysis of
line-profiles, as well as the examination of the CGM for galaxies
spanning a wide range of masses.  This will be the focus of our next
paper, which will consider whether the extended CGM of $z \sim 0$
galaxies can reproduce all of the observed statistics for \iovi\ as
argued by \cite{Prochaska2011}.

We further note that the \ihi\ surface
densities at $\rho < 50$\,kpc are sufficiently large that the CGM will
yield significant absorption from lower ionization states of heavy
metals (e.g.\ MgII, SiII, SiIII).  A proper estimate of the column
densities for these ions, however, will require a full treatment of
radiative transfer (i.e.\ to account for self-shielding by optically
thick \ihi\ gas).

The fact that the resultant extended, metal enriched CGM around our simulated galaxies matches the observations over a wide range of mass provides strong support for a vigorous baryon cycle in which outflows and subsequent cooling of halo gas play a key role in forming disc galaxies. 

\section*{Acknowledgements}
We thank Kate Rubin for useful conversations.
The analysis was performed using the pynbody package
(\texttt{http://code.google.com/p/pynbody}), which had key contributions from Rok Ro\v{s}kar
in addition to the authors.
The simulations were performed on the \textsc{theo} cluster of  the
Max-Planck-Institut f\"ur Astronomie at the Rechenzentrum in Garching;
the clusters hosted on \textsc{sharcnet}, part of ComputeCanada; the Universe
cluster that is part of the \textsc{cosmos} Consortium at Cambridge, UK; and the
\textsc{hpcavf} cluster at the University of Central Lancashire.  We greatly appreciate
the contributions of all these computing allocations.
CBB  acknowledges Max- Planck-Institut f\"ur Astronomie for its hospitality and financial support through the  Sonderforschungsbereich SFB 881 ``The Milky Way System''
(subproject A1) of the German Research Foundation (DFG).
J.X.P. is a research fellow of the Alexander
von Humboldt Foundation of Germany. HMPC and JW gratefully acknowledge the support of NSERC.  HMPC also appreciates the support he received from CIfAR.

\bibliography{references-1}

\clearpage

\end{document}